\documentclass[fleqn,usenatbib]{mnras}
\usepackage{bm}
\usepackage{newtxtext,newtxmath}
\usepackage{amsmath}
\usepackage{gensymb}
\usepackage{xcolor}
\usepackage{colortbl}

\usepackage[T1]{fontenc}

\DeclareRobustCommand{\VAN}[3]{#2}
\let\VANthebibliography\thebibliography
\def\thebibliography{\DeclareRobustCommand{\VAN}[3]{##3}\VANthebibliography}

\usepackage{graphicx}	
\usepackage{amsmath}	

\title[Restoration of contaminated observational data]{Restoration of contaminated data in an Intensity Mapping survey using deep neural networks}

\author[L. Li et al.]{
Lin-Cheng Li,$^{1}$
Jia-Yu Lin,$^{1}$
Yuan-Gen Wang$^{1}$\thanks{E-mail: wangyg@gzhu.edu.cn}
and Lister Staveley-Smith$^{2,3}$
\\
$^{1}$School of Computer Science and Cyber Engineering, Guangzhou University, 230 Waihuan West Road, Higher Education Mega Center, Guangzhou 510006, China\\
$^{2}$International Centre for Radio Astronomy Research (ICRAR), University of Western Australia, 35 Stirling Highway, Crawley, WA 6009, Australia\\
$^{3}$ARC Centre of Excellence for All Sky Astrophysics in 3 Dimensions (ASTRO 3D), Australia
}

\date{Accepted XXX. Received YYY; in original form ZZZ}

\pubyear{\the\year{}}

\begin{document}
\label{firstpage}
\pagerange{\pageref{firstpage}--\pageref{lastpage}}
\maketitle

\begin{abstract}
21-cm Intensity Mapping (IM) is a promising approach to detecting information about the large-scale structure beyond the local universe.
One of the biggest challenges for an IM observation is the foreground removal procedure. In this paper, we attempt to conduct restoration of contaminated data in an IM experiment with a Deep Neural Network (DNN). 
To investigate the impact of such data restoration, we compare the root-mean-square (RMS) of data with and without restoration after foreground removal using polynomial fitting, singular value decomposition, and independent component analysis, respectively.
We find that the DNN-based pipeline performs well in lowering the RMS level of data, especially for data with large contaminated fractions.
Furthermore, we investigate the impact of the restoration on the large-scale 21-cm signal in the simulation generated by CRIME. 
Simulation results show that the angular power spectrum curves from data with restoration are closer to the real one.
Our work demonstrates that the DNN-based data restoration approach significantly increases the signal-to-noise ratio compared with conventional ones, achieving excellent potential for IM observations.
\end{abstract}




\section{Introduction}
The statistical distribution of the matter distribution reflects the formation and evolution history of the universe and gives important insights into cosmological models.
Most observations to measure the large-scale distribution of baryonic matter are made at optical and infrared (IR) wavelengths
with ground-based and space telescopes  \citep{York00, Colless01, WigglZ, Euclid,  DESI, LSST, 2MASS, WISE}.
However, most wide-area optical/IR surveys have been limited to low redshifts and biased toward massive galaxies.

At radio wavelengths, 21-cm Intensity Mapping (IM) is a promising approach to measuring large-scale structure beyond the local universe across wide sky areas \citep{Chen15, Anderson18, CHIME2022, Zhang23}.
21-cm emission arises from the hyperfine transition line of neutral hydrogen (HI) and, due to the low bias of HI relative to dark matter \citep{Guha12}, it makes a good tracer for matter distribution.
21-cm observations over a broad frequency range also permit redshift measurements and allow the 3-D nature of large-scale structure and redshift space distortions to be traced. 

Studies using the HI IM technique are most easily applied in a multitracer framework where systematic errors can be mitigated by cross-correlation of the 21-cm signal with, for example, optical/IR data \citep{Pen08, Chang10, Masui2013, Anderson18, Visbal21,  Berti23}.
\cite{Pen08} reported the first detection of a cross-correlation signal between the HI Parkes All Sky Survey \citep[HIPASS,][]{Barnes01} and optical/IR data from the 6dF Galaxy Survey \citep[6dFGS,][]{Jones04, Jones09}. 
\cite{Chang10} used higher-redshift observations to detect a cosmological signal at redshift $z\sim$ 0.8 using data from the Green Bank Telescope (GBT) and WiggleZ \citep{WigglZ}.
\cite{Masui2013} and \cite{Switzer13} improved on this study. 
\cite{Anderson18} investigated the environmental dependency of HI content using the cross-correlation between Parkes and 2dFGRS strips across the North and South Galactic Poles.

However, except in the most radio-quiet sites, radio observations are often hampered by radio frequency interference (RFI) from intentional and unintentional emissions. Such interference can easily exceed the expected signal by many orders of magnitude. Additionally, Galactic and extragalactic foreground emission can also be many orders of magnitude larger than the expected HI temperature fluctuations, so small telescope chromaticity can easily dominate. This is particularly evident in Epoch of Reionization (EoR) experiments \citep{Parsons10, Paciga13, Van13, DeBoer17, Wayth18}.
Finally, the thermal noise from the telescope and receiver system will normally also be a few orders of magnitude higher than the HI signal. Therefore, the extraction of weak HI signals from observational data sets is a substantial challenge for IM surveys.

Mitigation of RFI is normally achieved by using the `cleanest' frequency ranges at the telescope site, and using a combination of automated or semi-automated flagging of data in the spectral and time domains \citep{Offringa10, Li2021, Sun22, Chakraborty22}.  
Foreground removal is conventionally achieved by using polynomial fitting in the spectral domain, which utilizes the spectral smoothness of the continuum emission \citep{Alonso15, Bigot15, Li2021}. However, the more successful foreground removal techniques use more advanced methods, including Singular Value Decomposition (SVD) and Principle Component Analysis (PCA) \citep{Davis85, Villaescusa17, Paciga11, Yohana21, JYE2023}, 
and Independent Component Analysis (ICA) \citep{Chapman12,Wolz14,Zhang16}.  
The principle idea for these methods is to utilize the different properties of the extragalactic HI signal and the foreground emission. 
Foregrounds are mainly diffuse synchrotron and free–free emissions, which are featureless along the frequency axis and, except for calibration errors, can theoretically be removed by smoothly varying functions. 
Similar techniques are used to deal with foreground estimation and removal in EoR experiments, where the HI signal is strong, but the synchrotron foreground is many times brighter than at higher frequencies  \citep{Wang06, Liu2011, Switzer13, Hothi21, WJY2021}.

Researches utilizing neural networks have achieved impressive outcomes in the fields of 21-cm IM.
\cite{Li19} firstly demonstrated that a convolutional denoising autoencoder can effectively extract EoR signal.
\cite{Makinen20} showed that a deep convolutional neural network is able to separate frequency and spatial patterns of HI from foregrounds and noise.
Based on these work, studies using neural networks have followed in different directions, including identifying ionized regions \citep{GH_21, Bianco21}, reconstruction of Laman-$\alpha$ maps \citep{Yoshiura22} and CO maps \citep{Zhou23}, elimination of beam effects \citep{Ni_22} and polarization leakages \citep{Gao23}.
These following studies have all obtained favorable results, demonstrating the powerful capability of neural networks.
For readers interested in the applications of machine learning in observational cosmology, we suggest \cite{Dvorkin22} and \cite{Moriwaki23} for a comprehensive review.

The works mentioned above are normally applied after the flagging of RFI since those signals can be orders of magnitude stronger than foreground emissions. This destroys all information
related to the cosmological and foreground signals at the contaminated frequencies and positions. However,
if a restoration can be made to the contaminated data, 
additional information may become available, assisting in better foreground removal.
In this paper, we therefore employ a convolutional neural network to restore contaminated data in an IM observation and investigate the impact of the data restoration in foreground removal.

The paper is organized as follows:
In Section 2, we describe the neural network and the loss function that we use in this paper.
Section 3 presents the observational data and the procedure that we extract the dataset for training.
In Section 4, we outline the foreground removal methods used in our analysis and present the restoration results.
Section 5 contains the ablation studies.
In Section 6, we establish a mock IM observation and apply our pipeline to analyze the impact on the detected large-scale signal.
Finally, Section 7 concludes our work. 

\section{METHOD}
\subsection{The neural network structure}
\label{sec:maths}  
The neural network that we adopt to restore the contaminated data is the inpainting network architecture - LaMa, proposed by \cite{Suvorov21}. 
In total, there are four architectures for LaMa: 1) LaMa-dilated using dilated convolutions in the convolution blocks; 2) LaMa-regular with regular convolutions in the convolution blocks; 3) LaMa-Fourier combining data in Fourier domain and space domain in convolution blocks; and 4) big-LaMa with doubled number of convolution layers in convolution blocks.
In \cite{Suvorov21}, LaMa-Fourier was found to achieve the best performance across a range of inpainting datasets at lower parameter and time costs than competitive baselines. 
\cite{Suvorov21} argue that it is because the LaMa-Fourier is good at restoring contaminated parts with periodic information such as nets, fences, and so on. 
However, our study finds that LaMa-dilated is a better alternative when the aim is to restore the foregrounds in contaminated areas (see below). 
In Figure~\ref{fig:architecture}, we illustrate the architecture of LaMa-dilated.
Details of the architectures of other LaMa models can be found in \cite{Suvorov21} and its supplementary document.

\begin{figure*} 
	\includegraphics[width=\textwidth]{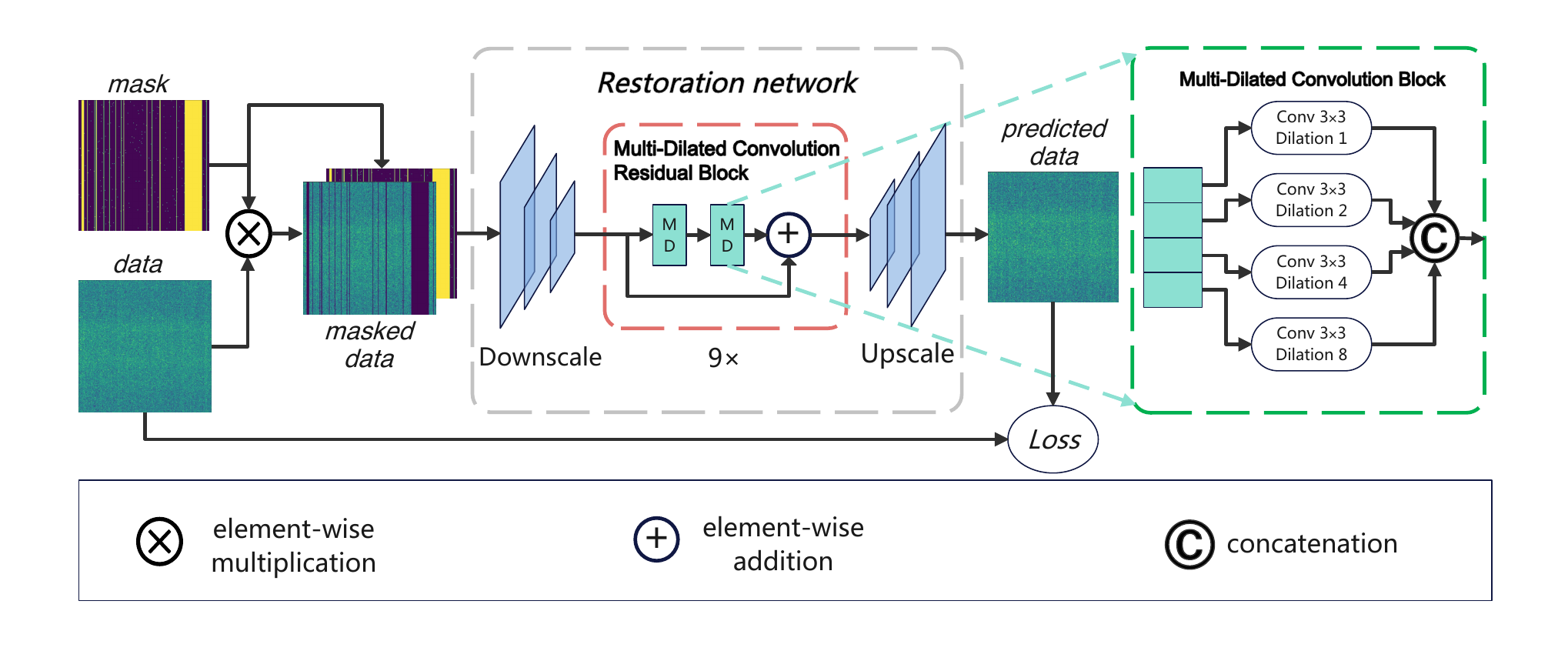}
    \caption{The architecture of LaMa-dilated neural network.}
    \label{fig:architecture}
\end{figure*} 

As shown in Figure~\ref{fig:architecture}, the LaMa-dilated network consists of three downsampling layers, a convolution block, and three upsampling layers.
The patch samples are firstly convolved with their masks and input along with their masks.
In the first block of the LaMa-dilated network, the input data are first downsampled three times in the downsampling layers, so as to reduce the data dimensions and allow the extraction of the global features in the following block.
The dilated convolution residual blocks then perform a series of dilated convolutions with different dilated lengths, which captures the features on different scales. 
The output of convolution blocks is finally convolved and upscaled to the original data dimensions, with the convolution kernel automatically trained.

\subsection{Loss function}
Following \cite{Suvorov21}, we use the loss function consisting of the high receptive field perceptual loss ($L_{\rm{HRFPL}}$), adversarial loss ($L_{\rm{adv}}$), and a feature matching loss.
We denote the input data, masks, and masked data as $x, m$, and $x \odot m$, respectively.
Then $L_{\rm{HRFPL}}$, which uses a high receptive field base model $\phi_{\rm{HRF}}(\odot)$, can be expressed as

\begin{equation} 
    L_{\rm{HRFPL}} = M([\phi_{\rm{HRF}}(x)-\phi_{\rm{HRF}}(\hat{x})]^2),
	\label{eq:quadratic}
\end{equation}
where $[\cdot-\cdot]$ is an element-wise subtraction operation, $M$ is the calculation operation on the sequence inter-layer mean of intra-layer mean, and $\phi_{\rm{HRF}}$ is a high receptive field base model.
The high receptive field perceptual loss focuses on the restoration from large masks and ensures the output quality of large-scale structures.

The adversarial loss is introduced to ensure that the generated data looks natural on small scales.
We denote the discriminator working on a patch level as $D(\cdot)$. 
The adversarial loss can then be expressed as

\begin{equation}
    L_{\rm{Adv}} = L_{\rm{D}} + L_{\rm{G}} + \lambda L_{\rm{GP}},
	\label{eq:loss_function}
\end{equation}

\begin{equation}
L_{\rm{D}} = -E_{\hat{I}}(\textrm{log}D(\hat{I})) -E_{\widetilde{I},m}[\textrm{log}D(\widetilde{I}\odot(1-m))]-E_{\widetilde{I},m}[\textrm{log}(1-D(\widetilde{I}\odot(1-m)))],
\end{equation}
and 
\begin{equation}
    L_{\rm{G}} = -E_{\widetilde{I}}[\textrm{log}D(\widetilde{I})],
\end{equation}
where $\hat{I}$ and $\widetilde{I}$ denote the ground-truth and predicted data, respectively, $L_{\rm{GP}} = E_{\hat{I}}\Vert\nabla_{\hat{I}}D(\hat{I})\Vert^2$ is the gradient penalty, and $\lambda = 1e^{-3}$.

Finally, the final loss also includes the gradient penalty $R_1 = E_x\Vert \nabla D(x)\Vert^2$, and the perceptual loss on the features of discriminator network ($L_{\rm{Disc}}$) which is used to stabilize training.

As a result, the final loss function for our model can be expressed as

\begin{equation}
    L_{\rm{final}} = \kappa L_{\rm{Adv}} + \alpha L_{\rm{HRFPL}} + \beta L_{\rm{Disc}} + \gamma R_1, 
    \label{eq:final_loss}
\end{equation}
where $\kappa=10$, $\alpha=30$, $\beta=100$, and $\gamma=0.001$ are hyper-parameters.  

\section{DATA}
\label{sec:data}
\subsection{Observation}
The observational data were obtained from the observing program P913 (PI: Staveley-Smith), taken with the CSIRO Parkes 64-m telescope in September and October 2016. 
This observation covers total $\sim$380 deg$^2$ in the frequency range above 700 MHz with a total of 60 hours of observing time.
During the observation a beamformer was used to form 17 discrete beams, each with a measured half-power width of 22.3$^{\prime}$ at the central frequency that we considered in this paper. 
Due to firmware limitations only 16 beams could be channelized. The beam spacing pitch was 0.35$^{\degree}$.
The surveyed regions were chosen to overlap the footprints of the 0h, 1h, 3h, and 22h fields of WiggleZ survey \cite{WigglZ} in order to allow a cross-correlation with optical density fields.
Depending on the RA of the targeted fields during the observation, the flux calibrators used for this observation were PKS B1934-638 and Hydra A.

To maximise the stability of the data, and to minimise the effect of sidelobe and RFI variation, the science observations were obtained in ‘shift-and-drift’ mode in which the azimuth and elevation of the telescope were fixed, and the telescope therefore scanned in RA at a fixed declination as the Earth rotated. 
At the frequency we consider here, RFI is mainly terrestrial in nature and there is more gain to be made by minimising exposure to RFI by drift scanning, than can be made in the reduction of 1/f noise by scanning the telescope at rates higher than sidereal. 
As demonstrated in \cite{Li2021}, the RFI in the observational data is relatively clean in the frequency range of 800 $\sim$ 820 MHz. 
We therefore use the data in this frequency range to apply our deep neural network. 
Table~\ref{tab:key_observational_parameters} lists the key parameters of the data used in this paper.

\begin{table}
	\centering
	\caption{Key parameters of the observational data used in this paper.}
	\label{tab:key_observational_parameters}
	\begin{tabular}{lccr} 
		\hline
		Parameter & Value\\
		\hline
		Bandwidth & 20 MHz\\
		Central frequency & 810.0 MHz\\
		Spectral resolution & 18.5 kHz\\
        Number of channels & 1080 \\
        Cycle time & 4.5 s \\
        Polarisations & 2 \\
        Beams & 16 \\
        Fields & 4 \\   
        Total integration time & 60 h \\
		\hline
	\end{tabular}
\end{table}
 
\subsection{Extraction of the training dataset} 
\label{sec:extraction}
\begin{figure} 
	\includegraphics[width=\columnwidth]{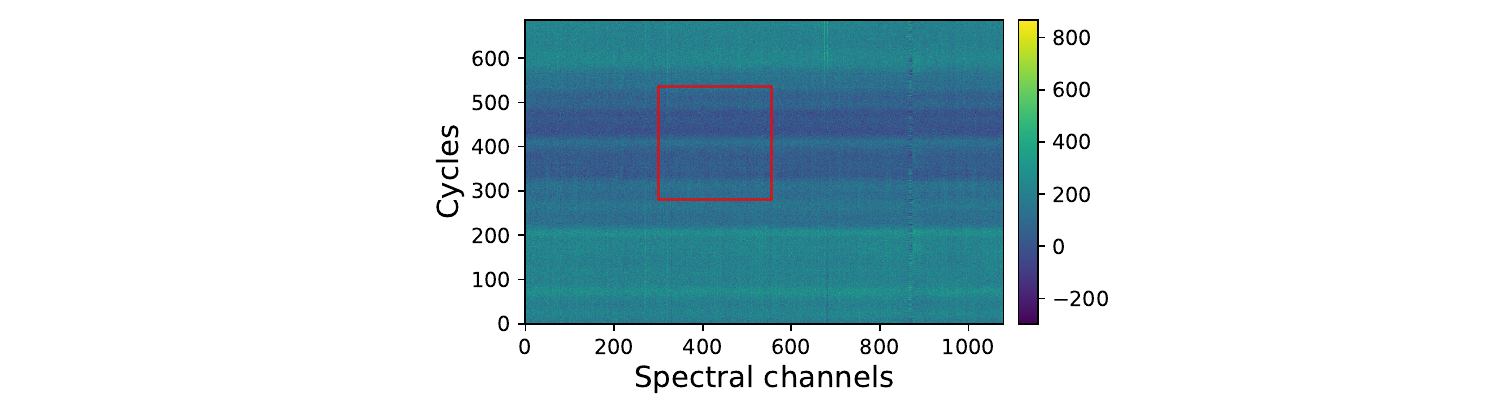}
    \caption{An example of the randomly extracted area (red square) from the original data.}
    \label{fig:patch_example}
\end{figure}

In order to generate the masks in training as real as possible, we utilize the masks in the observational data.
These masks were automatically generated during the processes of flagging RFI and foreground removal in \cite{Li2021}. 
Specifically, the channels with mean data values outside of the range $[\Bar{x}_1-3\sigma_1,\Bar{x}_1+3\sigma_1]$ e considered as the contaminated by RFI and are masked, where $\Bar{x}_1$ and $\sigma_1$ are the mean and standard deviation of data in each channel.
In each cycle, the data with values outside of the range $[\Bar{x}_2-3\sigma_2,\Bar{x}_2+3\sigma_2]$ are considered as the outliers contaminated by observational noise and are also masked, where $\Bar{x}_2$ and $\sigma_2$ are the mean and standard deviation of data in each cycle.
These positions of RFI and outliers are treated as the real mask for each patch sample of data.
To control the data quality in training, we set the threshold of masked fraction of each sample as 40$\%$.
Samples with a masked fraction higher than this threshold are considered as contaminated too much and are not included in our dataset. In each scan and each beam, we randomly extract square areas of data in the dimension of 256$\times$256 as the data patches.
In Figure~\ref{fig:patch_example}, we illustrate an example of one randomly selected patch overlaid on the raw data file.
To make full use of the raw data, each data file is sampled $n$ times for data patches depending on the dimensions of the data, with $n$ computed by 
$n = 3{n_1}n_{2}/S^2$,
where $S=256$ is the dimension of each sample, $n_1$ and $n_{2}$ are the dimension in cycles and the dimension in channels of the data, respectively.

In total, we extracted 23,806 patches of data, from which 5,000 patches are set as the validation set, and the rest are used as the training set.  
During the training, all models are trained for more than 3$\times10^5$ iterations until convergence with a batch size of 16 on NVIDIA GeForce RTX 4090 GPUs.

\begin{figure*} 
	\includegraphics[width=\textwidth]{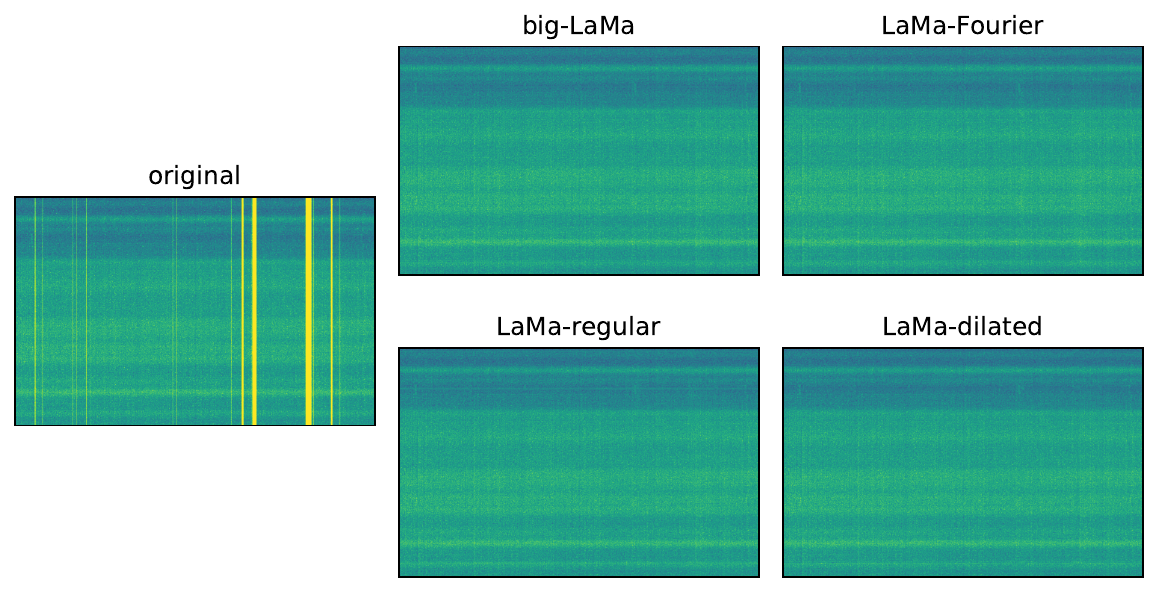}
    \caption{An example of the restoration results generated from different models.
    The leftmost panel is the original data with masks (yellow areas). 
    The other panels are the restoration results generated from big-LaMa, LaMa-Fourier, LaMa-regular, and LaMa-dilated, respectively, indicated by the titles of each panel.
    }
    \label{fig:restoration_example111}
\end{figure*} 

\subsection{Metrics}
\label{sec:metrics}
\begin{figure} 
	\includegraphics[width=\columnwidth]{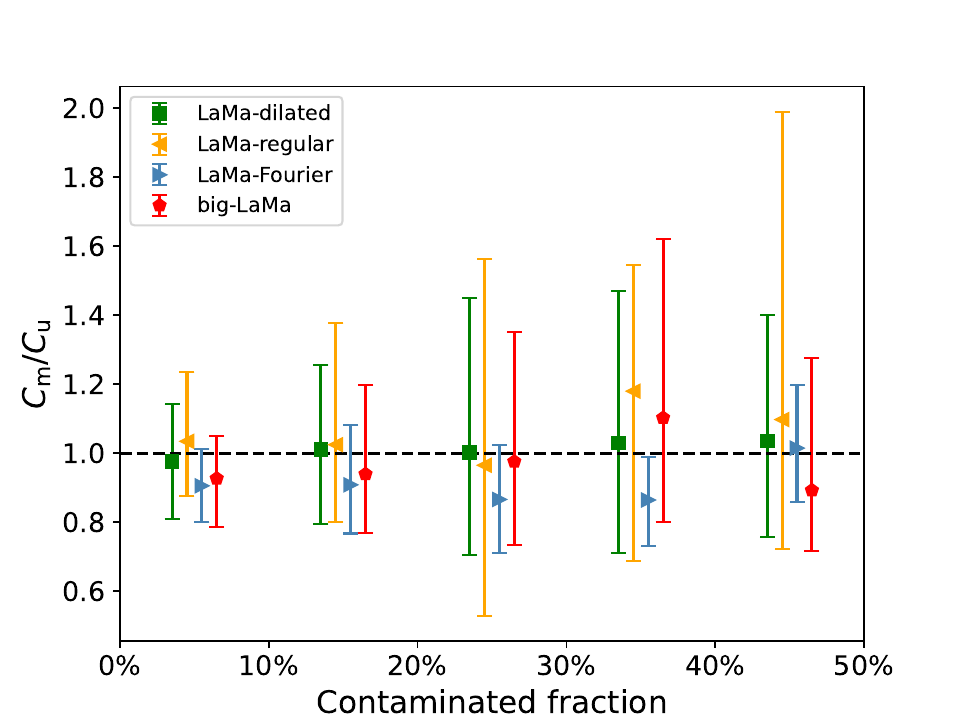}
    \caption{The median $C_{\rm{m}}/C_{\rm{u}}$ as the measurement of model performance for different models.
    Blue, orange, green, and red data points are the results for LaMa-dilated, LaMa-regular, LaMa-Fourier, and big-LaMa, respectively. Error bars represent the ranges from the 25th to the 75th percentiles in the prediction result of all samples in each interval. The coordinates of contaminated fractions for each model are shifted slightly for clarity.
    The dashed black horizontal line indicates the ideal value of 1.
    }
    \label{fig:ratio}
\end{figure}

In Figure~\ref{fig:restoration_example111} we show an example of the restoration result generated from different models.
The leftmost panel is the original data with masks (yellow areas). 
The other panels from left to right and top to bottom are the restoration results generated from big-LaMa, LaMa-Fourier, LaMa-regular, and LaMa-dilated, respectively.
As illustrated in Figure~\ref{fig:restoration_example111}, the restoration result from different models looks very similar making it difficult to visually decide which model outperforms the others.
We therefore use a metric related to the assumed foreground in masked areas in order to evaluate the performance of the different models. 
Considering that our restoration aims to recover the foregrounds as much as possible, we compute the ratio of $C_{\rm{m}}/C_{\rm{u}}$ as the performance metric,
where $C_{\rm{m}}$ and $C_{\rm{u}}$ are the element-wise cross-correlation values between foregrounds and the predicted data in the masked and unmasked areas, respectively. 
Because the predicted data are expected to be similar to the unmasked data, the ideal value of $C_{\rm{m}}/C_{\rm{u}}$ is 1.
Figure~\ref{fig:ratio} illustrates the values of $C_{\rm{m}}/C_{\rm{u}}$ in the prediction of all our samples for different LaMa models.
As shown in Figure~\ref{fig:ratio}, the values of $C_{\rm{m}}/C_{\rm{u}}$ for LaMa-dilated are the closest to 1 and the most stable when the contaminated fraction increases.
Therefore, we use the LaMa-dilated model to perform further analysis in the following sections.  
 
\section{Enhancement}
Various algorithms have been developed for the removal of foregrounds in the experiments of HI IM or detecting the EoR.
Polynomial fitting, SVD, and ICA methods are widely used and all have been demonstrated to be effective in the removal of foregrounds.
In this section, we investigate the impact of the restoration on foreground removal by comparing the results with and without restoration.
As RFI and foregrounds are typically several orders of magnitudes higher than HI fluctuations, a lower root-mean-square (RMS) level of data invariably means `cleaner' observational data with less contamination.
Therefore, we calculate the RMS of data after foreground removal as the measurement of data quality in the following analysis.

\subsection{Enhancement in polynomial fitting}
\begin{figure} 
	\includegraphics[width=\columnwidth]{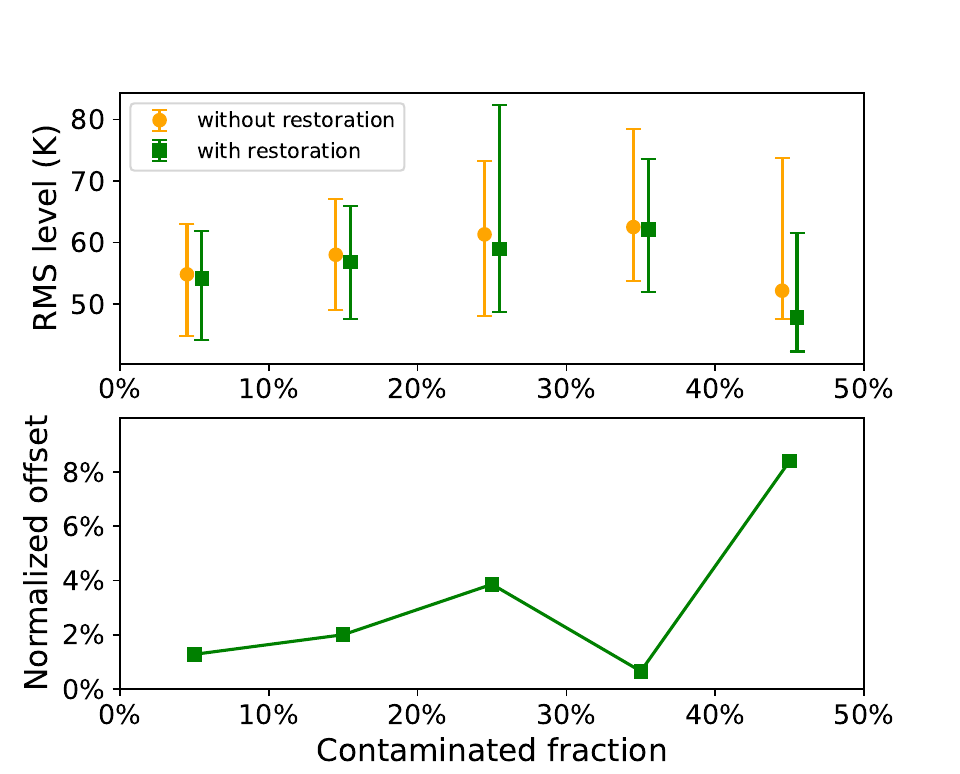}
    \caption{Top panel: The comparison of median RMS levels after foreground removal with second-order polynomials with (green) and without (orange) restoration. 
    Error bars indicate the ranges from the 25th to the 75th percentiles in each interval. 
    Bottom panel: The offsets between the points in the top panel, normalized by the values without restoration.
    }
    \label{fig:polyfit}
\end{figure}

Foreground signals are mainly made up of diffuse synchrotron and free-free emission, so their spectra are expected to be smooth along the frequency axis, unlike those of the HI signal.
As a result, subtraction of the foreground fitted with polynomials should permit a more accurate measurement of the cosmological HI fluctuations.   
In Figure~\ref{fig:polyfit} we compare RMS levels of the results after foreground removal with the subtraction of second-order polynomials between datasets with and without restoration.
The green and yellow points in the top panel are the RMS values with and without restoration, respectively.
The offsets between them are plotted in the bottom panel, normalized to the values without restoration.
As shown in Figure~\ref{fig:polyfit}, the RMS after restoration is lower than without restoration in all contaminated fraction bins.
The difference in RMS tends to increase with the increase in contaminated fraction. 
The results in Figure~\ref{fig:polyfit} confirm the effectiveness in Figure~\ref{fig:ratio} and demonstrate the robustness of our pipeline against large fractions of contamination.

\subsection{Enhancement in SVD}
\label{subsection:svd}
SVD is a generalization of standard eigenvalue decomposition techniques which can be applied to non-square matrices.
It decomposes the data cube into orthogonal modes of decreasing amplitude, with the low-order modes mainly containing the unwanted foreground and RFI residuals. 
Generally, the 2-D HI data matrix in dimensions of $m\times n$ is decomposed in an SVD process by  

\begin{equation} 
    M = U\Sigma V^*,
	\label{eq:svd}
\end{equation}
where $U$ is the $m\times n$ left-vector matrix, $\Sigma$ is the $m\times n$ matrix where the singular values are the non-negative reals on the diagonal, and $V^*$ is the conjugate transpose of $V$, which is the $n\times n$ matrix of right-singular vectors.
The singular values of $M$ represent the amplitudes of each mode after decomposition. 
As the amplitudes of interference are assumed to be higher than the large-scale HI signal in an IM observation, the low-order SVD modes are expected to mainly consist of this interference, which facilitates a ‘cleaner’ detection of large-scale signals once these modes are removed and the HI data matrix is reconstructed from $M$.

\begin{figure} 
\includegraphics[width=\columnwidth]{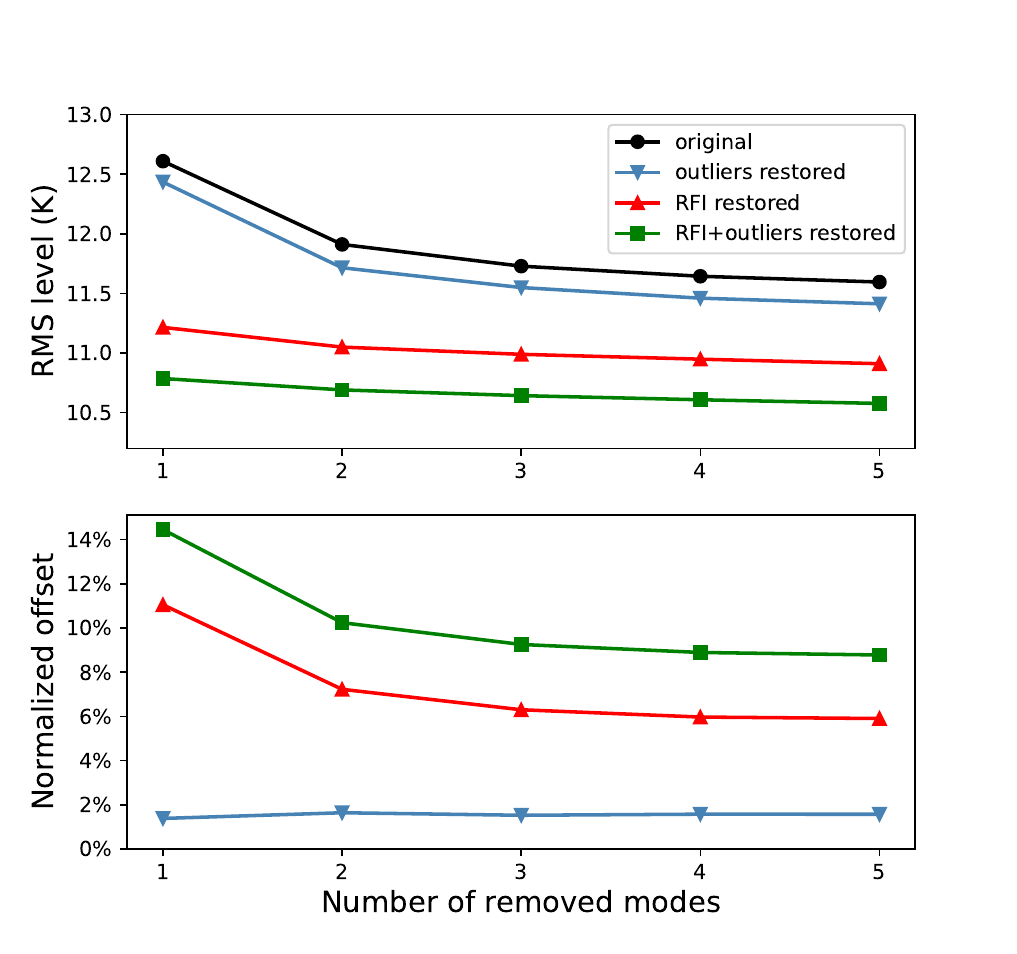}
    \caption{Top panel: the standard deviation of different datasets as functions of a number of SVD modes removed.
Curves from top to bottom are the median standard deviation results in datasets containing the original data, restoration for outliers, restoration in RFI channels, and restoration in positions of both RFI channels and scattered outliers, respectively.
Bottom panel: the normalized offset between each line to the black line in the top panel.}
    \label{fig:svds}
\end{figure}

However, the number of modes that should be removed from the HI matrix depends on the properties of the interference.
As demonstrated in \cite{Li2021}, the modes of large-scale HI signal are coupled to the overall fluctuation. 
Thus, we investigate the effect of our restoration pipeline using different numbers of removed modes.
In the investigation, we generate four datasets based on how much the data are restored from our pipeline:
a) the original dataset without restoration;
b) the dataset with restoration at positions of outliers;
c) the dataset with restoration of RFI channels; and
d) the fully restored dataset with restoration in the position of RFI channels and outliers.
In Figure~\ref{fig:svds} we plot the RMS for datasets with different numbers of modes removed via SVD.
As previously discussed, a higher RMS value indicates that the datasets are more affected by interference.
As shown in Figure~\ref{fig:svds}, the RMS level of original datasets with RFI is the highest among the four datasets. 
In contrast, the green curve of datasets with restoration for both scattered outliers and RFI channels is the lowest, indicating that our pipeline can effectively reduce the noise after the foreground removal in SVD.
 
\subsection{Enhancement in ICA}
Since the early 1980s, ICA has been a technique to blindly separate 
individual signals from their mixture, using the assumption that the signals to be separated are independent of each other.
\cite{Hyvarien99} proposed FastICA as an improved version of ICA with much higher calculation efficiency. 
This method has been applied in the foreground removal by \cite{Maino02}, \cite{Bottino10}, \cite{Chapman12}, and \cite{Wolz14}.
A comprehensive introduction and tutorial can be found online\footnote{ \url{http://cis.legacy.ics.tkk.fi/aapo/papers/IJCNN99_tutorialweb/}}.  
We briefly outline its basic principle here.

The equation of ICA is usually expressed as 
\begin{equation}
    \mathbf{x} = \mathbf{\rm{A}s} = \sum_{i=1}^{n}{a_i}{s_i}, 
\end{equation}
where $\mathbf{x}$ is the observed signal, $\mathbf{\rm{A}}$ is the mixing matrix to be calculated, $\mathbf{s}$ contains the independent components (ICs) to be separated, and $n$ is the number of assumed ICs. 
In practice, FastICA separates components in all lines-of-sight simultaneously.
The dimensions of $\mathbf{x}$, $\mathbf{\rm{A}}$, and $\mathbf{s}$ are (number of measurements)$\times$ (number of frequency channels)$, $(number of measurements)$\times n$, and $n\times$ (number of frequency channels), respectively.  
To solve equation 8, $\mathbf{s}$ can be formulated as 
\begin{equation}
    \mathbf{s} = \mathbf{\rm{A}^{-1}x} = \mathbf{Wx} , 
\end{equation}
where $W$ is the inverse of $\mathbf{\rm{A}}$.

\begin{figure} \includegraphics[width=\columnwidth]{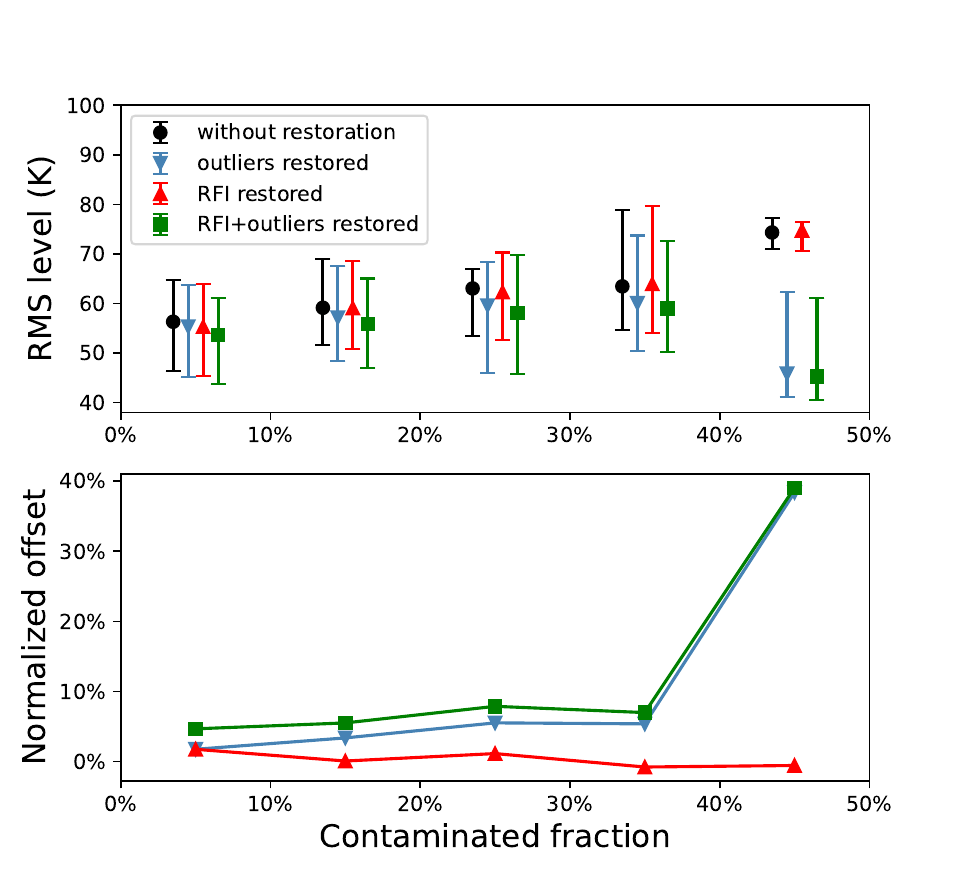}
    \caption{Top panel: the RMS levels after foreground removal with FastICA in different datasets. The error bars represent the ranges from the 25th to the 75th percentiles in each interval.
    Bottom panel: the normalized offsets to the values of the dataset without restoration.
     }
    \label{fig:ica}
\end{figure}

According to the central limit theorem, the distribution of the sum of independent variables is more Gaussian than individual components.
So FastICA seeks components by minimizing the Gaussianity after the separation.
To measure the Gaussianity, FastICA calculates the negentropy based on the idea that the entropy of a Gaussian distribution is the largest among the other distributions.
The negentropy of variable $y$ is calculated via  
\begin{equation}
    J(y)=H(y_{\rm{gauss}})-H(y) , 
\end{equation}
where $H(y)=-\int p(y)\log y\, {\rm d}y$, and $y_{\rm{gauss}}$ is a Gaussian variable with the same covariance matrix as $y$.
Considering that the calculation of $H(y)$ is costly, the kurtosis variable is often used in practice as the measurement of Gaussianity:

\begin{equation}
    kurt(y) = E(y^4) - 3E(y^2)^2 . 
\end{equation}
Based on the FastICA algorithm, we separate four components from our datasets.
After subtracting a few components from the datacube, the residuals are expected to be the combination of HI signal, noise, and foreground fitting errors.
In Figure~\ref{fig:ica} we compare the RMS levels of the residuals after the removal of four ICs in the four datasets used in section~\ref{subsection:svd}.
As can be seen in Figure~\ref{fig:ica}, the RMS level of d) dataset (green) is the lowest in each bin, indicating that the restoration also helps to suppress interference in foreground removal with FastICA. 

\section{Ablation studies}
\label{sec:ablation}
To investigate the role of different parts of our method, we perform ablation studies with some parts of our model changed or removed. 
Specifically, the changes include: 
a) some losses are removed from the training process;
b) different dimensions of patch samples in the training set; 
and c) different metrics.

\subsection{Different losses}
Here we examine the role of components in the loss function by removing losses from the final loss and calculate $C_{\rm{m}}/C_{\rm{u}}$ again for the new loss function.
According to Equation~\ref{eq:final_loss}, the components that we considered include $L_{\rm{Adv}}$, $L_{\rm{Disc}}$, $L_{\rm{HRFPL}}$, and $R_1$.

We found that the training process cannot converge and get corrupted if $L_{\rm{Adv}}$ or $R_1$ is removed from $L_{\rm{final}}$, which indicates that the model relies on $L_{\rm{Adv}}$ and $R_1$ to work effectively and stable.
When $L_{\rm{Disc}}$ is deleted, $C_{\rm{m}}/C_{\rm{u}}$ after data restoration reaches 1.021.
While the $C_{\rm{m}}/C_{\rm{u}}$ becomes 0.976 if only $L_{\rm{HRFPL}}$ is removed in LaMa-dilated.
As we introduced in subsection~\ref{sec:metrics}, the ideal value of $C_{\rm{m}}/C_{\rm{u}}$ is 1, so the results of $C_{\rm{m}}/C_{\rm{u}}$ have become worse 
compared to the result using original loss where none component is removed, demonstrating that both $L_{\rm{disc}}$ and $L_{\rm{HRFPL}}$ benefit the performance of our model.
In Table~\ref{tab:ablation} we summarise the results of $C_{\rm{m}}/C_{\rm{u}}$ of the ablation study.

\begin{table}
	\centering
	\caption{The ablation results of $C_{\rm{m}}/C_{\rm{u}}$ if different components are removed from $L_{\rm{final}}$ in the ablation study. The symbol of '$\backslash$' in the result column means that the neural network does not converge or corrupted in training if the corresponding component is removed.}
	\label{tab:ablation}
	\begin{tabular}{cc} 
		\hline
		Removed components & $C_{\rm{m}}/C_{\rm{u}}$\\
		\hline
		$L_{\rm{Adv}}$ & $\backslash$ \\
		$L_{\rm{HRFPL}}$ & 0.976\\
		$L_{\rm{Disc}}$ & 1.021\\
        $R_1$ &  $\backslash$ \\
        None & 1.001 \\ 
		\hline
	\end{tabular} 
\end{table}

\begin{figure} 
	\includegraphics[width=\columnwidth]{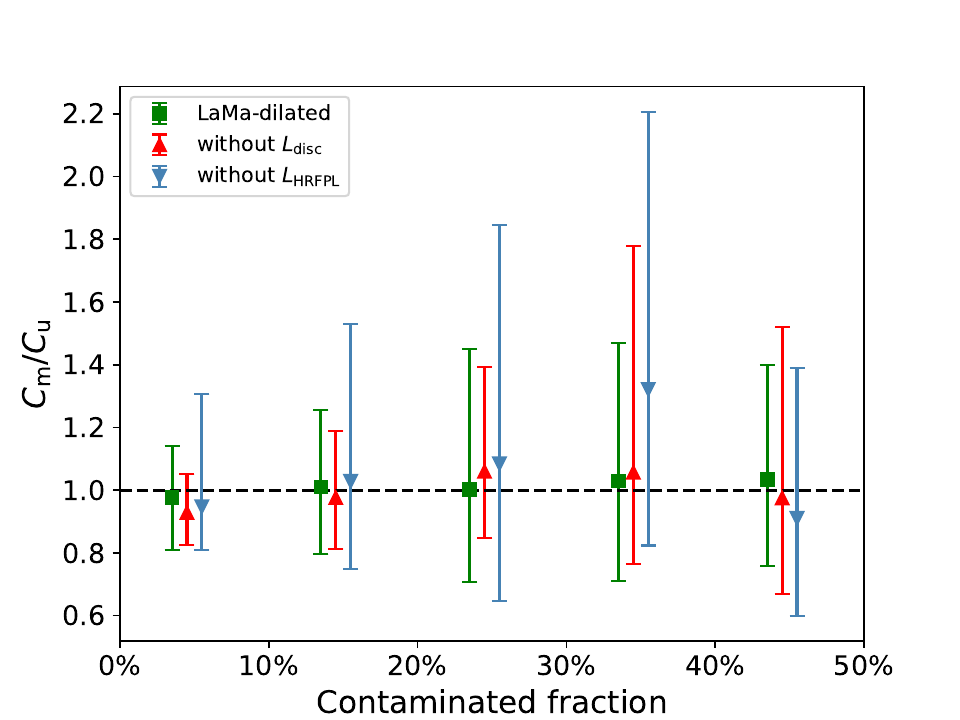}
    \caption{The values of $C_{\rm{m}}/C_{\rm{u}}$ for models using different loss functions.
    The blue, yellow, and green points are the results of the LaMa-dilated model, LaMa-dilated model without $L_{\rm{disc}}$, and without $L_{\rm{HRFPL}}$, respectively. 
    The error bars represent the ranges from the 25th to the 75th percentiles in each interval. 
    The black horizontal dashed line indicates the ideal value of 1.
     }
    \label{fig:cc_losses}
\end{figure}

\subsection{Training sets with different dimensions} 
\begin{figure*} 
	\includegraphics[width=\textwidth]{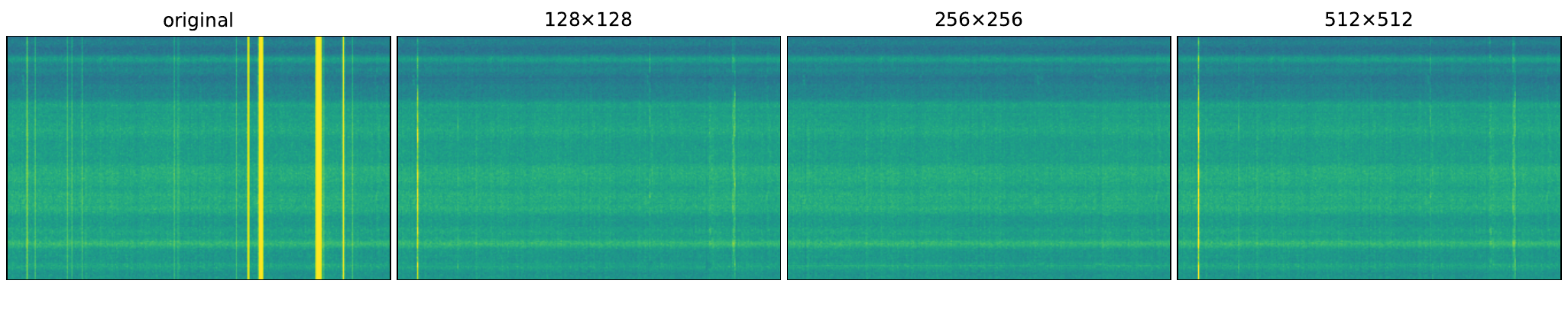}
    \caption{An example of restoration results generated by models trained with samples of different dimensions.
    Panels from left to right are the original data with masks (yellow areas), the restoration results generated from models trained with samples of dimension 128$\times$128, 256$\times$256, and 512$\times$512, respectively. 
     }
    \label{fig:dimensions}
\end{figure*}

Intuitively, training sets of images with larger dimensions may help the model to capture more large-scale information in the raw data.
Because the telescope was set in a drift-scan mode, the dimension along the time axis depends on the cycle time and total integration time for each scan. 
Some datasets have dimensions less than 512 and thereby cannot be sampled to generate patches of 512$\times$512 -- the other dimension being frequency.
This results in smaller sizes for the training set which may degrade the training process.
On the contrary, the strategy of using small patches of data 
can help to increase the number of training samples but make it hard to capture the large-scale information of samples during training.
Here we investigate the impact of sample dimension by comparing the restoration results using models trained with samples of 128$\times$128 ($M_{128}$), 256$\times$256 ($M_{256}$) and 512$\times$512 dimensions ($M_{512}$), respectively. 
The $M_{128}$ and $M_{512}$ training sets includes 116,534 and 3,919 samples, respectively.
Figure~\ref{fig:dimensions} shows an example of data restored from models trained with patches of different dimensions.
The leftmost panel is the original data with masks (yellow areas).
The other panels from left to right are the restoration results generated from $M_{128}$, $M_{256}$, and $M_{256}$, respectively.
The restoration results generated from $M_{128}$ look worse than for $M_{256}$, indicating more difficulty for $M_{128}$ to glean large-scale information from samples of a smaller dimension during the training process, even though the size of training set is a few times larger.
The restoration using $M_{512}$ also looks worse than for $M_{256}$ despite the extra dimension size. 
This seems due to the small size of the training set. 

\subsection{Different metrics}
We note that the performance of LaMa-dilated using the $C_{\rm{m}}/C_{\rm{u}}$ metric surpasses the LaMa-Fourier models while the latter is the best model in \cite{Suvorov21}. 
To investigate the reason for this,
we calculate the $SSIM$ (Structure Similarity Index Measure) and $PSNR$ (Peak Signal-to-Noise Ratio), which are two indices widely used in the field of image-quality-assessment. 
The $SSIM$ between two images of $x$ and $y$ is computed using 

\begin{equation} 
    SSIM(x,y) = \frac{(2\mu_x\mu_y+c_1)(2\sigma_{xy}+c_2)}{(\mu^2_x+\mu_y^2+c_1)(\sigma_x^2+\sigma_y^2+c_2)},
	\label{eq:ssim}
\end{equation}
where: $\mu_x$ and $\mu_y$ are the mean pixel values of $x$ and $y$, respectively; $\sigma_x$ and $\sigma_y$ are the standard deviations of $x$ and $y$, respectively; $\sigma_{xy}$ is the covariance between $x$ and $y$; $c_1=(k_1L)^2$ and $c_2=(k_2L)^2$ are parameters to keep the fraction stable with $L$ equal to the dynamic range of pixels, $k_1=0.01$ and $k_2=0.03$.

$PSNR$ is defined by

\begin{equation} 
    PSNR = 10\times \rm{log}_{10}(\frac{MAX^2}{MSE})=20\times \rm{log}_{10}(\frac{MAX}{\sqrt{MSE}}),
	\label{eq:psnr}
\end{equation}
where $\rm{MAX}$ is the maximum pixel value and $\rm{MSE}$ is the mean square error between the predicted data and the original data.
Higher values of $SSIM$ or $PSNR$ conventionally indicate better restoration quality. 

In the work of \cite{Li2021}, the foregrounds are fitted with polynomials and subtracted from the observational data for each integration cycle. 
We take the foregrounds extracted in the pipeline of \cite{Li2021} as the ground-truth of our prediction in the calculation of $SSIM$ and $PSNR$. 
Table~\ref{tab:metrics} lists the results of $SSIM$ and $PSNR$ for the four models which are considered in this work.
As shown in Table~\ref{tab:metrics}, $SSIM$ and $PSNR$ of LaMa-Fourier are the best of all models, which is consistent with the results of \cite{Suvorov21}.
Therefore, we argue that it is due to the use of different metrics that LaMa-dilated is the best model evaluated in our study. 
In our work, we adopt $C_{\rm{m}}/C_{\rm{u}}$ as the metric since we require the model to better emphasize correlation over texture in the restored areas.
In contrast, $SSIM$ and $PSNR$ give more weight to the formation of similar textures compared to unmasked areas in order to generate natural-looking images.
Considering that the aim of restoration in this study is to recover the contaminated foregrounds and to facilitate further data analysis, we argue that $C_{\rm{m}}/C_{\rm{u}}$ is a more relevant metric than $SSIM$ and $PSNR$, and that LaMa-dilated is the optimal model. 

\begin{table}
	\centering
	\caption{The performance of different models in datasets.}
	\label{tab:metrics}
	\begin{tabular}{lccr} 
		\hline     
		models & $SSIM$ & $PSNR$ \\
		\hline 
        LaMa-dilated & 0.939 & 36.945 \\
        LaMa-Fourier  & 0.944 & 37.971 \\
        LaMa-regular  & 0.932 & 36.355 \\
        big-LaMa      & 0.937 & 36.438 \\
		\hline
	\end{tabular}
\end{table}

\section{Simulation}
The key scientific goal of an IM experiment is to measure the large-scale information of neutral hydrogen.
Thus, a natural question rises that whether the finally measured large-scale signal may be affected by the data restoration since the restored data may cause new contamination in following processes of foreground removal. 
To figure out this, we establish a mock IM survey utilizing the CRIME\footnote{\url{http://intensitymapping.physics.ox.ac.uk/CRIME.html}} simulation code \citep{Alonso14}, which is a public software for generating fast simulations of intensity mapping observations.
In the simulations, the main components of the signals include: cosmological 21cm signal, galactic free-free emission, galactic synchrotron, extra-galactic free-free emission, and point sources radiation.
Because our goal in this section is to analyze how the data restoration affects the measured large-scale signal in a mock observation, we do not introduce observational noise and beam effect in the simulation for clarity.
In the following subsections, we describe how the main components in the simulations are modeled.

\subsection{Cosmological HI signal}
The cosmological HI signal is usually expressed as the brightness temperature $T_b$.
In the Rayleigh-Jeans limit ($hv_{21}\ll k_BT_b)$, the brightness temperature can be written as 
\begin{equation} 
    T_b(\boldsymbol{\hat{n}},\nu) = \frac{3c^3A_{21}(1+z)^2}{16k_B\nu_{21}^2H(z)}n_{\textrm{HI}}(z,\boldsymbol{\hat{n}}),
    \label{Tbequation}
\end{equation}
where $\boldsymbol{\hat{n}}$ 
is the line of sight, $\nu$ is the frequency, 
$\hbar$ is the reduced Plank's constant, $c$ is the light speed, $n_{\rm{HI}}$ is the comoving number density of HI, $k_B$ is the Bolzmann's constant,
$H(z)$ is the Hubble parameter as a function of redshift $z$, 
and $\nu_{21}$ is the frequency of the HI emission line at $z$=0.
In a standard flat $\rm{\Lambda}$CDM model, 
$T_b$ can be related to the HI overdensity 
$\delta_{
\rm{HI}}=\rho_{\rm{HI}}/\overline{\rho}_{\rm{HI}}-1$ with
\begin{equation} 
    T_b(\boldsymbol{\hat{n}},z) = 190.55\frac{\Omega_bh(1+z)^2x_{\rm{HI}}(z)}{\sqrt{\Omega_m(1+z)^3+\Omega_{\rm{\Lambda}}}}(1+\delta_{\rm{HI}}(\boldsymbol{\hat{n}},z)) \ \rm{mK},
\end{equation}
where $h=H_0/(100km/s/Mpc)$ is the dimensionless Hubble constant, $x_{\rm{HI}}$ stands for the neutral hydrogen mass fraction relative to the total baryons, and 
$\Omega_b$, $\Omega_m$, $\Omega_{\rm{\Lambda}}$ correspond to the present fractions of baryon, total matter, and dark energy density, respectively.
The CRIME code firstly generate a Gaussian overdensity field along with the velocity potential.
Then the overdensity field and radial velocity is derived to each cell.
The total HI mass in each cell thereby can be calculated as 
\begin{equation}
    M_{\rm{HI}} = 2.775\times10^{11}M_{\odot}V_c\frac{\Omega_bx_{HI(z)}}{h}(1+\delta_{\rm{HI}}),
\end{equation}
where the $M_{\odot}$ is the solar mass, and $V_c$ is the volume of each cell.
Finally, the brightness temperature calculated through equation~\ref{Tbequation} is projected onto the sky map using the HEALPix pixelization scheme.

\subsection{Foregrounds}
\begin{figure}
    \centering
	\includegraphics[width=0.88\columnwidth]{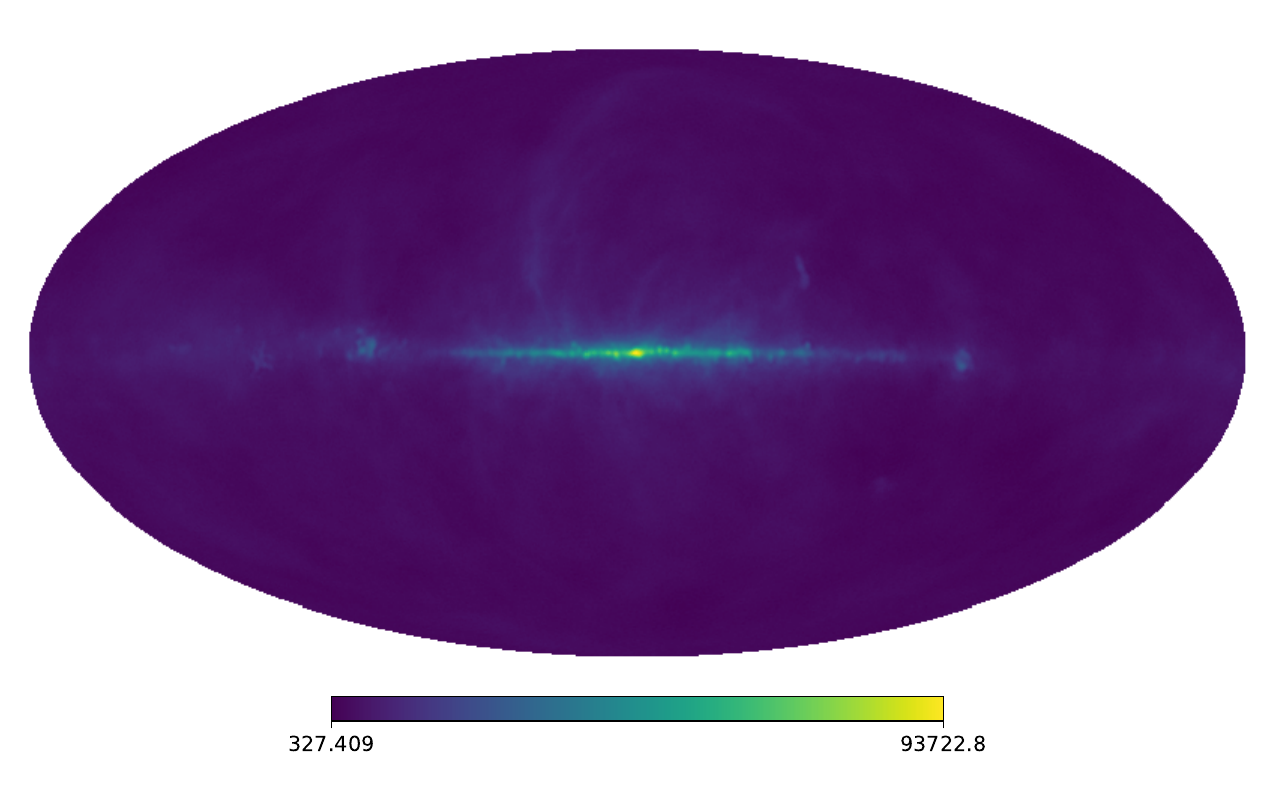}
    \caption{The total intensity of the HealPix map generated by CRIME in the spectral channel of 810 MHz, in unit of mK.}
    \label{fig:skymap}
\end{figure}

The foregrounds in an IM observation typically consists of 5 components: polarised and unpolarized galactic synchrotron, galactic and extragalactic free-free emission, and the emission from extragalactic and point radio sources. 
The CRIME code assumes isotropic extragalactic point sources and free-free foregrounds.
Following \cite{Santos05}, their power spectra are modeled as 
\begin{equation}
C_l(\nu_1,\nu_2)=A\left(\frac{l_{\rm{ref}}}{l}\right)^\beta\left(\frac{\nu_{\rm{ref}}}{\nu_1\nu_2}\right)^\alpha \rm{exp}\left(-\frac{\rm{log}^2(\nu_1/\nu_2)}{2\xi^2}\right),
\end{equation}
where $\xi$ is the frequency frequency-space correlation length of the emission, which regulates its spectral smoothness.
The parameters used for the different foregrounds were taken from \cite{Santos05} and are listed in Table~\ref{tab:alphabeta}.
Galactic synchrotron is yielded by cosmic ray electrons interacting with galactic magnetic field. 
From Table~\ref{tab:alphabeta} we can see that the galactic synchrotron is the most dominant in the foreground emission.
Considering this, the CRIME code separates this radiation as polarized and unpolarized emission and model these two kinds of emission in a careful manner.
We refer the reader to \cite{Alonso14} for a comprehensive description of their model.
Figure~\ref{fig:skymap} demonstrates the total intensity of the HealPix map of radiations generated using CRIME in the central spectral channel of 810 MHz.

\begin{table}
	\centering
	\caption{Foreground $C_l(\nu_1,\nu_2)$ model from \citet{Santos05} for the pivot values $l_{\rm{ref}}=1000$ and $\nu_{\rm{ref}}=130$ MHz.}
	\label{tab:key_parameter}
	\begin{tabular}{ccccc} 
		\hline
		Foreground & A(mK)$^2$ & $\beta$ & $\alpha$ & $\xi$  \\
		\hline
		Galactic synchrotron & 700 & 2.4  & 2.80 & 4.0 \\
		Point sources & 57  & 1.1 &  2.07 & 1.0 \\
		Galactic free-free & 0.088 & 3.0 & 2.15 & 35 \\
        Extragalactic free-free & 0.014 & 1.0 & 2.10  & 35 \\
		\hline
	\end{tabular}
\label{tab:alphabeta}
\end{table}

\subsection{Generating data sets in mock observation}
To align with our real observational data as much as possible, the frequency range of the mock IM observational data generated with CRIME code are also set in the frequency range from 800$\sim$820 MHz with the same spectral resolution of 18.5 kHz.
The coordinates of the mock observation are arbitrary set as $20^{\circ}<ra<50^{\circ}$, and $25^{\circ}<dec<55^{\circ}$, in the dimension of 512$\times$512.
Further, we randomly pick the RFI sample in real observation to match the data in the mock observation as their RFI signals.
Data in mock observation located in flagged positions are added with the corresponding RFI signal. 
Then in the same way as Section \ref{sec:data}, we extract data patches which are in dimensions of 256$\times$256. 

In total, 11500 data patches are extracted from the mock observational data.
Among them, 2000 data are set as validation set and the rest are set as the training set.
With the mock data sets, the Lama-dilate network is trained and then make data restoration to the mock observational data.

\subsection{Difference in foreground removal}
\begin{figure*} 
	\includegraphics[width=\textwidth]{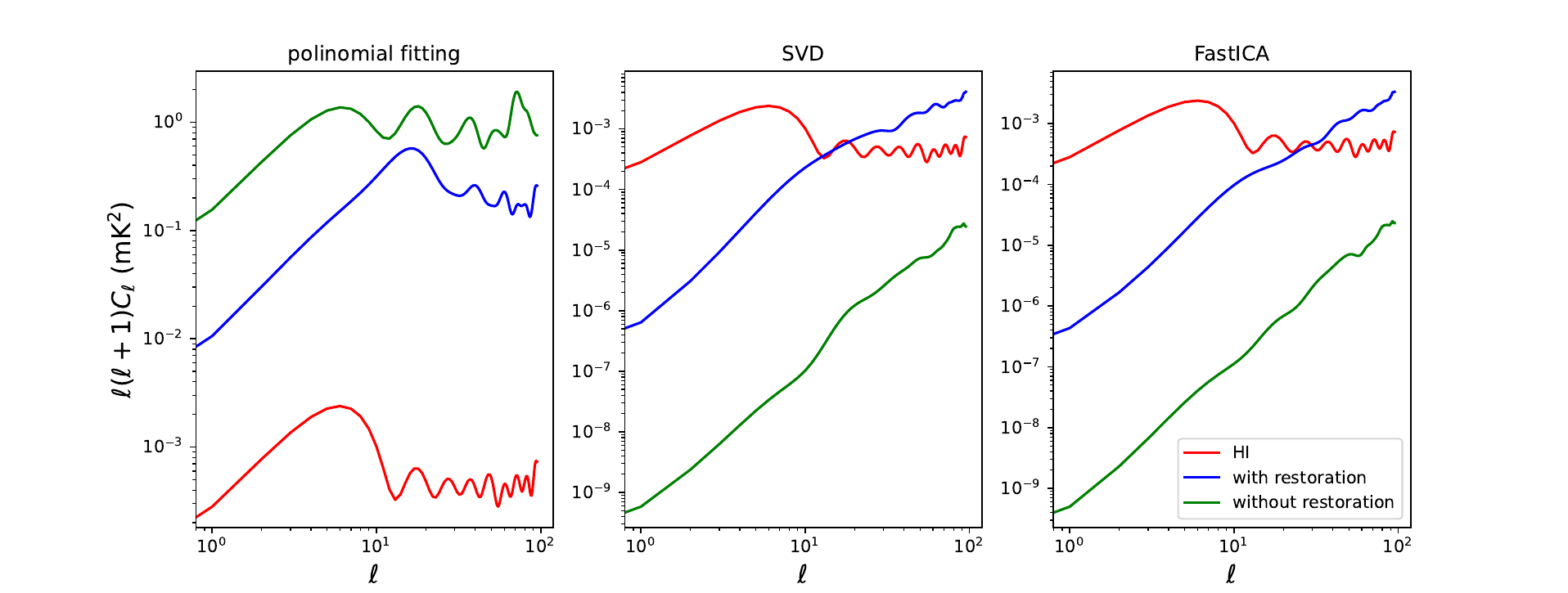}
    \caption{Angular power spectra of different data after foreground removal using different methods. 
    Blue, and green curves represent the results for mock observational data with and without restoration, respectively.
    For comparison, the angular power spectrum of real HI signal is also plotted as the fiducial curve in red.
    The foreground removal method performed in each panel is denoted as the title of each subplot. 
    }
    \label{fig:draw3ps}
\end{figure*}

To figure out whether the data restoration affects the final observed large-scale signals,
in this section, we make tests of 2-order polynomial fitting, SVD, and FastICA using the mock data with and without data restoration.
To speed up the calculation in the following anylasis, we downsampled the our data along the frequency axis by averaging every 20 spectral channels as one channel.
Due to some broad-band RFI, there are no valid data in some channels after masking the data.
Because the foreground removal methods cannot deal with data with empty values, we fill these empty channels with the mean values in their corresponding cycles.
Then the filled data are performed with foreground removal methods, and the angular power spectrum ($Cl$) of the residual is calculated.
Figure\ref{fig:draw3ps} shows the results of angular power spectra of different data performed with different foreground removal methods.
When conducting SVD and FastICA, we subtract four components from the original datacubes.
Blue, and green curves represent the results for mock observational data with and without restoration, respectively.
For comparison, the angular power spectrum of real HI signal is also plotted as the fiducial curve in red.
The method used for foreground removal is denoted as the title in each subplot.
As can be seen in Figure~\ref{fig:draw3ps}, the angular power spectrum of data set with data restoration is closer to the fiducial values in log space compared to the result without restoration.
Specifically, results of polynomial fitting method show that the angular power spectra curve of data with restoration is higher than the pure HI signal, but is lower than the data without restoration.
In the tests of SVD and FastICA, the angular power spectrum curve also has been moved towards the fiducial curve of pure HI after restoration is implemented. 
We argue that this difference is because the data restoration helps to enhance the data quality in contaminated channels, which makes the data better prepared for the foreground removal.

\section{Conclusion}
21-cm IM observations suffer contamination from RFI. 
Data restoration to the contaminated areas may provide more information to the pipeline and benefit following analysis.
In this paper, we have made attempt to conduct restoration of contaminated data in an IM experiment using a deep neural network. The observational data are obtained from an IM survey covering $\sim380$ deg$^2$ of the sky in a frequency range from 800 to 820 MHz using 60 hrs of observing time. We generated samples of patches from the data, which were used to train the model.
By comparing the results using a cross-correlation metric $C_{\rm{m}}/C_{\rm{u}}$, the LaMa-dilated model is considered as the best model.

We have investigated the impact of restoration on foreground removal using polynomial fitting, SVD, and FastICA.
For each of the foreground removal methods, our results show that restoration helps to suppress the noise level, especially for data with larger contaminated fractions. 
An ablation study on the loss function, the dimensions of the training set, and the metrics have been also conducted.


To further investigate the impact of our pipeline on the observed large-scale HI signal,
we have utilized the CRIME code to generate a mock IM observation. 
Results from the mock observation show that the angular power spectrum curve has been moved towards the fiducial values after data restoration, especially in the low $l$ end.
This result indicates that the large-scale HI signal suffers less distortions in the pipeline with data restoration compared to that without restoration, and that the restoration helps to enhance data quality for foreground removal.

Our study has explored the possibility of restoring contaminated radio astronomy data in order to detect large-scale structure, which highlights the potential of increasing the signal-to-noise ratio of the observational data through the pre-processing using data restoration.
With future large datasets from the SKA and other instruments, it will be useful to explore the restoration of contaminated data with more sophisticated neural networks.

\section*{Data Availability}

The observational data used in this study are not fully public but may be obtained by contacting Lister Starveley-Smith at lister.staveley-smith@uwa.edu.au with a detailed request, which will be evaluated based on scientific merit and data usage compliance.



\bibliographystyle{mnras}
\bibliography{main} 





\bsp	
\label{lastpage}
\end{document}